\documentclass[twocolumn,showpacs,preprintnumbers,amsmath,amssymb,9pt]{revtex4}

\usepackage{graphicx}
\usepackage{dcolumn}
\usepackage{bm}

\newcommand{\be}{\begin{equation}}
\newcommand{\ee}{\end{equation}}
\newcommand{\ba}{\begin{eqnarray}}
\newcommand{\ea}{\end{eqnarray}}

\def\be{\begin{equation}}
\def\ee{\end{equation}}
\def\barr{\begin{array}}
\def\earr{\end{array}}
\def\bea{\begin{eqnarray}}
\def\eea{\end{eqnarray}}

\def\to{\rightarrow}
\def\be{\begin{equation}}
\def\ee{\end{equation}}
\def\bea{\begin{eqnarray}}
\def\eea{\end{eqnarray}}

\def\sF{{{\rm F}\!\!\!\!\hskip.8pt\hbox{\raise1pt\hbox{/}}\,}}

\def\r{\rho}

\def\x{\xi}

\begin{document}
\preprint{hep-th/0611021}
\title{A New Phase at Finite Quark Density from AdS/CFT}
\author{Shin Nakamura,$^{a,b}$ Yunseok Seo,$^{a}$
Sang-Jin Sin,$^{a}$ and K. P. Yogendran, $^{a,b}$}
\vskip 0.5cm
\affiliation{$^a\,$ Department of physics,BK21 Program Division,
Hanyang University, Seoul 133-791, Korea\\
$^b \,$Center for Quantum Spacetime, Sogang University,
Seoul, 121-742, Korea
}
\date{\today}

\begin{abstract}
We explore phases of an ${\cal N}=2$ super
Yang-Mills theory at finite quark density by introducing
quark chemical potential in a D3-D7 setup.
We formulate the thermodynamics of brane embeddings
and we find that the density versus chemical potential
equation of state has rich structure.
This yields {\em two} distinct
first order phase transitions in a small window of quark density.
In order words, there is a new first order phase transition
in the region of deconfined quarks. In this new phase, the chemical potential is a decreasing function of the density. 
We suggest that this might be relevant to the difference in
sQGP--wQGP phases of QCD.
\end{abstract}

\pacs{Valid PACS appear here}
\maketitle

{\bf Introduction: }
There has been much hope that one might be able to use AdS/CFT
\cite{AdS/CFT} to describe the real systems after certain amount of
deformations. For example, it has been suggested that the fireball in
Relativistic Heavy Ion Collision (RHIC) viewed as a
strongly interacting system \cite{RHIC-1,RHIC-2},
has been studied using dual gravity models
\cite{son,SZ,Nastase,yaffe,jns}.
There have been many attempts to construct models 
phenomenologically closer to QCD
\cite{AdS/QCD}.

More recently, there has been renewed interest
in ${\cal N}=2$ super Yang-Mills (SYM) systems with quenched
fundamental
quark flavors studied by using a holographic description 
with probe D7-branes in the $AdS_{5}$ black hole background
\cite{Myers6,johnson,kk,myers1,myers2,evans}.
The key observation is that we have confinement of quarks
even in the absence of gluon confinement or area law \cite{Myers6}.
The phases of this theory are characterized by the brane embeddings:
whether the D7-brane touches the black hole horizon 
(black hole embedding) or not
(Minkowski embedding).
Different types of embedding lead to different meson spectra.

\begin{figure}[!ht]
\begin{center}
{\includegraphics[angle=0,
width=0.3\textwidth]{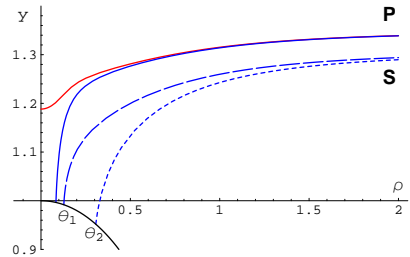} }
\caption{\label{fig:embed} Brane embedding and  phase transitions. Solid lines:
Minkowski embedding (red) to black hole embedding (blue).
Dashed lines: Hopping from a black hole embedding to another one.}
\end{center}
\end{figure}

In this letter, we explore the phases of
this theory 
at {\it finite quark density}
by introducing quark chemical potential
along the lines of \cite{horigometani,kimsinzahed}.
We will first establish a clear formulation of
the thermodynamics of brane embeddings.
We find that we need to renormalize the
finite chemical potential due to
the divergence of the thermodynamic potentials.
We will also find that apart from the type of
first order phase transition described 
in \cite{Myers6,johnson} at zero chemical potential,
there is another class of first order phase transition
within the black hole embedding category: it is indicated in
Fig.~\ref{fig:embed} as a hopping between two black hole embeddings.

Since black hole embeddings correspond to
the deconfined phase,
we cautiously suggest that this new type of first order phase 
transitions
might be relevant to the difference between sQGP--wQGP 
in RHIC experiments. In particular, we find that the chemical potential in this new phase is a decreasing function of the density. 

We emphasize that
depending on whether we control the system by the chemical potential
(grand canonical ensemble) or by the density (canonical ensemble),
the phase diagram is different.
In this letter,
we present the analysis and the results for the system based 
on the canonical ensemble.
The details including the description
of the other process based on the grand canonical ensemble will be
reported in detailed publication \cite{work2}.

{\bf The Euclidean AdS black hole metric} is given by
\begin{eqnarray}
ds^{2}
= \frac{U^{2}}{R^{2}}\left(f(U) dt^{2}+d\vec{x}^{2}\right)+
R^{2}\left( \frac{dU^{2}}{f(U) U^{2}}+d\Omega_{5}^{2}\right),
\label{adsm}\end{eqnarray}
where $f(U) =1-\left( {U_{0}}/{U}\right)^{4}$.
The Hawking temperature of this geometry is given by
$T= {U_{0}}/{\pi R^{2}}= {U_{0}}/({\sqrt{2\lambda}\pi\alpha'})$
where $\lambda=g_{YM}^{2}N_{c}$.
We introduce a dimensionless coordinate $\xi$ defined by
$ {d\xi^2}/{\xi^2}= {dU^2}/{(U^2 f)}$
, so that
the bulk geometry is
\ba
ds^{2}
&=&\frac{U^{2}(\xi)}{R^{2}}
\left(f(\xi) dt^{2}+d\vec{x}^{2}\right)
+
\frac{R^{2}}{\xi^{2}}ds^2_{6}, \quad {\rm with}~  \\
ds^2_{6}&=& d\xi^{2}+\xi^{2}d\Omega_{5}^{2}=
 d\rho^{2}+\rho^{2}d\Omega_{3}^{2}+dy^{2}
+y^{2}d\varphi^2 ,\nonumber
\ea
where we have defined
$\xi^{2}\equiv y^{2}+\rho^{2}$ and 
$\rho$ is the radius of the 3-sphere.
The black hole horizon is located at $\xi=1$.
The {\it induced metric} on the D7-brane is
\begin{eqnarray}
ds_{D7}^{2}
=
\frac{U^{2}}{R^{2}}\left(f dt^{2}+d\vec{x}^{2}\right)
+
\frac{R^{2}}{\x^2}
\left((1+y'^{2})
d\rho^{2}+\rho^{2}d\Omega_{3}^{2}\right),\label{indm}
\end{eqnarray}
where $y'=\partial_{\rho}y(\rho)$.
It is interesting to notice that the bulk metric (\ref{adsm})
and the induced metric (\ref{indm}) have
the same Hawking temperature.
This means that the bulk and the brane are in equilibrium.
The Euclidean DBI action of the D7-brane in the presence of
the gauge field strength $F_{\rho t}$ is%
\begin{eqnarray}
S\!=\!N_{f}\mu_{7}\!
\int dt d^{3}x d\rho d\Omega_{3}
\sqrt{{\rm det}(G+2\pi\alpha' F)}
\!=\!\beta V_{3}\int d\rho {\cal L}
, 
\end{eqnarray}
with
\begin{eqnarray}
{\cal L}&=& \tau_7 
 {\rho}^{3}\omega_{+}^{3/2}
\sqrt{\frac{\omega_{-}^{2}}{\omega_{+}}
(1+ {y}'^{2})-(F/m_{T})^{2}},  \label{L}
\end{eqnarray}
where $\tau_{7}=N_{f}N_{c}^{2}T^{4}g_{YM}^{2}/32$,
$\omega_{\pm}(\x) =1\pm {\xi}^{-4}$ and
$m_T=\frac{1}{2}\sqrt{\lambda}T$
(the Chern-Simons term vanishes in the
present case).

{\bf Conserved charge and equations of motion: }
Since ${\cal L}$ does not depend on $A_0$ explicitly,
its conjugate momentum is a {\em conserved quantity}:
\begin{eqnarray}
\Pi_{A_0}=\frac{\partial {\cal L}}{\partial F_{\rho t}}
\equiv -Q ,
\label{consv}
\end{eqnarray}
in terms of which we can write
\begin{eqnarray}
F_{\rho t} =
\:m_{T}
 {\tilde{Q}\omega_{-}\sqrt{(1+ {y}'^{2})}
 }/ {\sqrt{\omega_{+} (\tilde{Q}^{2}+\omega_{+}^{3}  {\rho}^{6})}},
\label{F-tilde}
\end{eqnarray}
where $\tilde{Q}=\frac{m_{T}}{\tau_{7}}Q$.
Since we have a constraint (\ref{consv}), 
to obtain a Lagrangian for $y$, we should not substitute 
$F_{\r t}$ into the original Lagrangian (\ref{L}).
The correct procedure is to perform a Legendre transformation
\begin{eqnarray}
{\cal H}= {\cal  L}- \Pi_{A_0}{A'_0}, \label{H}
\end{eqnarray}
and then impose the conservation equation
(\ref{consv}) to eliminate the electric field completely.
The resulting ``Hamiltonian''  is given by
\begin{eqnarray}
{\cal H}
={\cal T}( {y}, {\rho})
\sqrt{1+ {y}'^{2}},\:\:
{\cal T}( {y}, {\rho})=\tau_{7}
\sqrt{\frac{\omega_{-}^{2}}{\omega_{+}}
(\tilde{Q}^{2}+\omega_{+}^{3}{\rho}^{6})}.
\end{eqnarray}
We can take this Hamiltonian  as our
effective Lagrangian for ${y}$ and ${\cal T}$ may be regarded as
the effective tension of the D7-brane. Resulting
equation of motion  is
\begin{eqnarray}
\frac{ {y}''}{1+ {y}'^{2}}
+\frac{\partial \log {\cal T}}{\partial  {\rho}} {y}'
-\frac{\partial \log {\cal T}}{\partial  {y}}=0.
\label{Eomy}
\end{eqnarray}
One can check that both the original and the effective
Lagrangian give the same equation of motion for $y$.

{\bf Chemical potential in gravity dual: }
In Refs. \cite{kimsinzahed,horigometani},
the quark chemical potential was introduced
as the  value of $A_0$ on the D7-brane worldvolume.
Here we define the chemical potential in a gauge-invariant fashion:
\begin{eqnarray}
\mu  \equiv
\int^{\infty}_{ {\rho}_{min}}d {\rho}\:F_{\rho t}
=\lim_{ {\rho}\to\infty}A_{0}. \label{chem-def}
\end{eqnarray}
For the last equality, we need to gauge fix $A_\rho=0$ and set
$A_{0}|_{ {\rho}_{min}}=0$, which
agrees with \cite{kimsinzahed,horigometani}.

Notice that
$\mu = \int F_{\rho t}$ is the work
to bring a unit charge from the UV region ($\rho=\infty$)
to the IR region against electric field $F_{\r t}$.
This definition (\ref{chem-def}) agrees with our
intuition of the chemical potential as work done to
add a quark to the system.

For the Minkowski embedding, ${\rho}_{min}=0$.
For the black hole embedding, ${\rho}_{min}=\cos\theta$
where $\theta$ is the angular coordinate on the $y$-$\rho$ plane.

{\bf Thermodynamic potentials:}
A generic grand potential (density) is defined by
$e^{-\beta V_{3}\Omega(\mu)}={\rm Tr}{e^{-\beta V_{3}(H-\mu N)}}$.
Here we identify the DBI action which is a functional of $A_0$
as the grand potential 
$S_{DBI}=\beta V_{3}\Omega$.
Then, integrating the Legendre transformation (\ref{H}),
\begin{eqnarray}
\int d\rho { \cal H} =  \int d\rho {\cal L} -
\int d\rho \Pi_{A_0} {A_0'}.
\end{eqnarray}
Using the fact that $\Pi_{A_0}=-Q$ is a constant
(in $\rho$ evolution),
one can rewrite the above as
\begin{eqnarray}
F(Q)=\Omega(\mu)+\mu Q.
\end{eqnarray}
It is remarkable that the Legendre transformation in the
bulk classical field theory is reinterpreted as the Legendre
transformation between the canonical and the grand canonical
ensembles in the boundary thermodynamics.

The chemical potential
enters the Hamiltonian density of the gauge theory
at the boundary as a coupling to the baryon number
density:
\be
\Delta {\cal H}_{YM}=- \mu\langle\psi^\dagger \psi\rangle.
\ee
Therefore $Q$ which has been originally
defined as a first integral of the DBI action
should be identified as the number density of quarks/baryons.
Notice that the effective tension of the brane increases
as we increase the quark density which may have been expected.
More precisely, after considering various scale factors, we have
\begin{eqnarray}
Q=
\langle\psi^\dagger \psi\rangle.
\label{tau}
\end{eqnarray}

{\bf Renormalization of chemical potential:}
The Helmholtz free energy $F$ and
the grand potential are in fact divergent quantities
since they contain a divergent $\rho$ integral.
Therefore we need to regularize them.
We choose to subtract $F_0(Q)$,
the value of $F$ for the D7-brane configuration
that touches the black hole on the equatorial plane
($y=0$).
This is like a Pauli-Villars regularization in the brane setup.
We call this brane as the reference brane.
So the renormalized free energy which we will calculate
is defined by
\be
F_R(Q)=F(Q)-F_0(Q).
\ee
This has a
far reaching effect to the chemical potential.
To see this notice
\begin{eqnarray}
F_R(Q)\!=\!\Omega(\mu)\!-\!\Omega(\mu_0)
\!+\!(\mu\!-\!\mu_0) Q
\!:=\!\Omega_R(\mu_R) \!+\! \mu_R Q,
\end{eqnarray}
where we have used the fact that
$F_0(Q)=\Omega(\mu_0)+\mu_0 Q$.
Notice that $\mu_0$ is the chemical potential at
the reference-brane configuration.
The point is that when we deal with the renormalized quantities
like $F_R$ and $\Omega_R$, we also have to use the 
renormalized chemical potential,
although $\mu$ itself is finite quantity from the beginning.
In  all numerical analysis,
we need to use $\mu_R$ for the chemical potential.

From now on, we delete subindex $R$
unless it is confusing.

{\bf The phase structure of D3-D7:}
In this article, we study the system
based on the canonical ensemble, where
the number density $Q$ is continuous while the chemical potential is
allowed to jump.

Let us expand $y$ and $A_{0}$ in the form of
${y}({\rho})=L+\frac{\tilde{c}}{{\rho}^{2}}+O({\rho}^{-4})$,
${A_{0}}({\rho})=\mu+\frac{c'}{{\rho}^{2}}+O({\rho}^{-4})$.
We have $L=\frac{m_{q}}{\sqrt{2\lambda}\pi T}$, $ \tilde{c}=c\frac{\sqrt{2}\pi m_{T}}{\tau_{7}}  $
where $m_{q}$ is a quark mass and $c=
\langle\bar\psi\psi\rangle$
 \cite{evans,myers2}.
One can interpret $L$ as the quark mass at
fixed temperature or as inverse temperature at fixed quark mass.
We also find that $c'=-\frac{1}{2}\frac{m_{T}^{2}}{\tau_{7}}Q$.
The standard AdS/CFT dictionary
establishes that
we have two pairs of conjugate variables, namely  $(\tilde{c},L)$ and
$(Q, \mu)$.

\begin{figure}[!ht]
\begin{center}
{\includegraphics[angle=0,
width=0.3\textwidth]{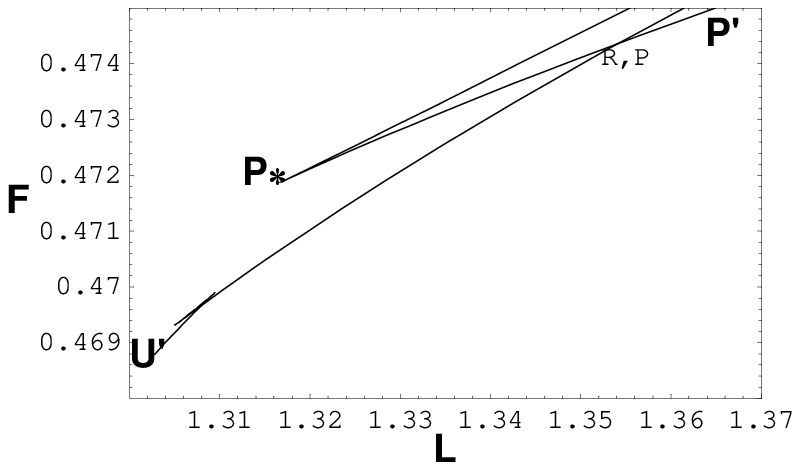} }
\caption{\label{fig:FL} Free energy vs. $L$ for fixed $Q$.
The second phase transition is indicated by small kink near U'.}
\end{center}
\end{figure}

To analyse the phase transition, we plot the free energy
$F/T^{4}$ as a function of $L$ for a given $Q$
(Fig.~\ref{fig:FL}) and track the least free energy
configuration.
In Figure ~\ref{fig:FL}, we start from P' on the Minkowski branch. As $L$ decreases, the free energy decreases until we intersect the black hole branch at P.
For smaller values of $L$, the black hole configurations have lower free energy.  At P therefore, 
the D7-brane jumps from a Minkowski embedding
to touch the black hole horizon.
This is the same type of first order phase transition first
found in \cite{Myers6} for zero chemical potential.
Decreasing $L$ further, decreases the latitude ($\theta$) 
of the intersection
of the brane with the horizon smoothly. 
For large $Q$, the story ends here. 

\begin{figure}[!ht]
\begin{center}
{\includegraphics[angle=0,
width=0.3\textwidth]{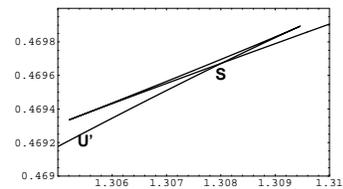} }
\caption{\label{fig:FL2} Free energy vs. $L$ for fixed $Q$
around the second phase transition point.}
\end{center}
\end{figure}

\begin{figure}[!ht]
\begin{center}
{\includegraphics[angle=0,width=0.3 \textwidth]{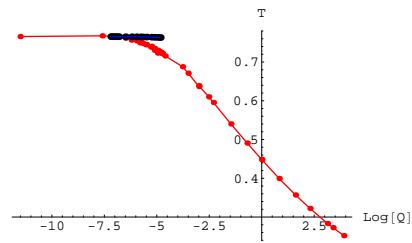} }
\caption{ \label{fig:TQ} Phase diagram in $T$-$\tilde{Q}$ plane.}
\end{center}
\end{figure}

For very small $Q$, however, as we decrease $L$ further, initially
the latitude of intersection of brane-black hole goes down smoothly.
But, at a critical latitude $\theta_1$,
the embedding suddenly jumps
to a smaller latitude $\theta_2$ as indicated schematically 
in Fig.~\ref{fig:embed} (the kink near U' in Fig.~\ref{fig:FL}).
An embedding with touching latitudes between
these two values has higher free energy as seen
in Fig.~\ref{fig:FL2}, and hence is never realized.
This phenomenon happens only at finite density within a
small density window,
\begin{eqnarray}
\log\tilde{Q}_{1*}
=-6.812< \log\tilde{Q}<-4.726
=\log\tilde{Q}_{2*}.
\end{eqnarray}
The full phase diagram therefore looks as in Fig. \ref{fig:TQ}.
Notice that the horizontal axis is $\log\tilde{Q}$.
The chemical potential jumps across the phase boundary.

\begin{figure}[!ht]
\begin{center}
{\includegraphics[angle=0,width=0.3 \textwidth]{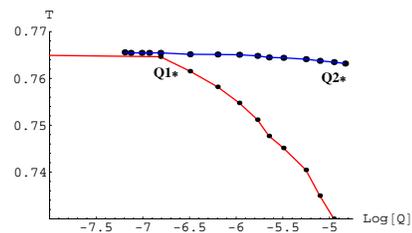} }
\caption{ \label{fig:TQ2} $T$-$\tilde{Q}$ phase diagram around
the new phase boundary.}
\end{center}
\end{figure}

For $\tilde{Q}> \tilde{Q}_{2*}$,
there is no kink and hence the second phase transition disappears
as we can see in Fig. \ref{fig:TQ2}. Furthermore,
we have a second order phase transition at $\tilde{Q}= \tilde{Q}_{2*}$.
The slopes of the phase boundary lines are discontinuous at 
$\tilde{Q}=\tilde{Q}_{1*}$, where three lines meet.

{\bf Equations of state: }
The relations between the variables $(\tilde{c},L)$ (equivalently $(c,m_q)$) and
$(Q, \mu)$
can be considered
as the diagram of equation of state, which
is much like the $P$-$V$ diagram of Van der Waals
in liquid-gas phase transition.
We may determine these by using the thermodynamic relations
$$c=-\frac{\partial F}{\partial m_q} , \hspace{1cm} \mu=\frac{\partial F}{\partial Q}.$$

Numerical analysis shows that
the relationship between $\mu$ and $Q$ given by
$Q=-\frac{\partial \Omega(\mu)}{\partial \mu}$
is not monotonic, further $\mu$ is not a single-valued function
of $Q$.

\begin{figure}[!ht]
\begin{center}
{\includegraphics[angle=0,
width=0.3\textwidth]{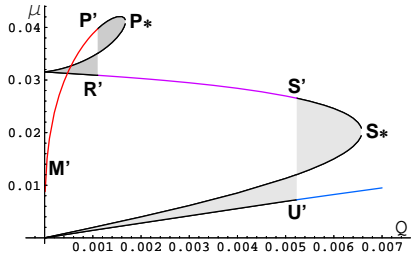} }
\caption{ \label{fig:muQ} $\mu$-$Q$ relation: We plot $\mu/T$
vs. $Q/T^{3}$.}
\end{center}
\end{figure}

\begin{figure}[!ht]
\begin{center}
{\includegraphics[angle=0,
width=0.3\textwidth]{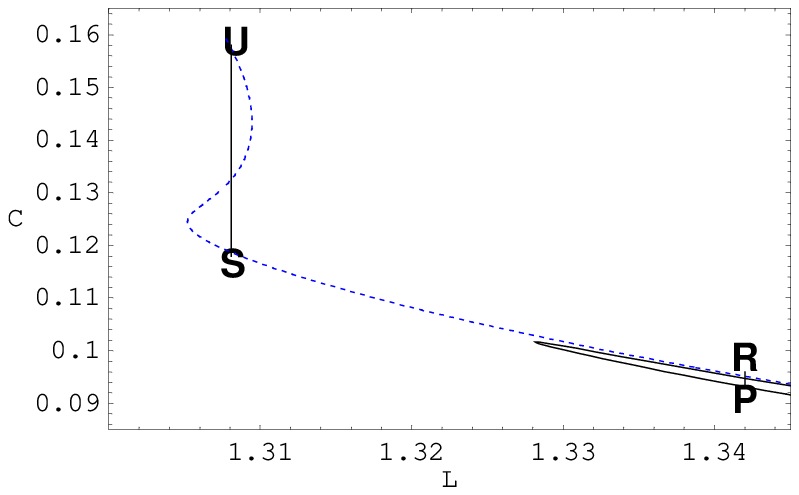}}
\caption{  \label{fig:cL} $\tilde{c}$-$L$ relation: Here we interpret $L$ as quark mass.}
\end{center}
\end{figure}

In Fig.~\ref{fig:muQ} and \ref{fig:cL}, we show the relation 
between $\tilde{c}$ and $L$ for a representative value of $Q$ and
the relation between $Q$ and $\mu$ for a representative 
value of $L$.
The equations of state represented in these diagrams show that
we have  much richer structure than Van der Waals $P$-$V$ diagram.

We track the phase diagram from M' in Fig. \ref{fig:muQ} at a fixed value of $L$ which may be thought of inverse temperature 
(or quark mass). 
The first phase transition takes place
at a certain critical value of $Q$ at point P'.
Here, the brane embedding jumps from a Minkowski 
embedding to a black hole embedding which results in 
a jump of the chemical potential from P' to R' on the diagram.
If we increase $Q$ further, we have a second phase
transition which is realized as a jump from S' to U'. 
We have also indicated the Maxwell's construction, which allows us
to determine the location of the phase transition (i.e., equality of the area of the shaded region on either side of a particular jump).
For larger values of $L$ however, we have only a single phase transition. 

The point to be noted is that in a small window between
R' and S', the chemical potential is a {\em decreasing} function of the density (in the deconfined  phase). 
One way to achieve this is to have bound states. 
If this is true, our new phase is likely to have many bounded pairs 
of quarks.

Referring to the $\tilde{c}$-$L$ equation of state in Fig.~\ref{fig:cL}, 
the two phase transitions are as labelled. In this case however, 
the chiral condensate decreases uniformly as a function of 
$L$ (excepting of course, for the jumps).
For large quark mass $L$, it seems to
vanish as expected.

%

{\bf Discussion:}
Although characterized by nonstandard behaviour of the chemical potential, the nature of the second phase is not very clear 
from the gauge theory point of view. However,
since both phases 
belong to the black hole
embedding corresponding to deconfined quarks,
it might be relevant to the famous difference between
sQGP and wQGP discussed within RHIC physics
\cite{RHIC-1}. 
In fact if we take the temperature and $\tilde{Q}$
of the relevant region to be 200MeV and $5\times10^{-3}$ respectively and choose $ {N_{f} N_{c}\sqrt{\lambda}}/{16} \sim O(1)$, then the
density $Q\sim 1 \:\:{\rm fm}^{-3}$,
which is in the RHIC ballpark. 

An immediate question is the universality of this second phase. Since most of the properties of the branes, for small $Q$, are determined in the neighborhood of the horizon, we might expect that this phenomenon persists in (perhaps more realistic) models so long as we have a black hole.



One can also look at a process which corresponds to a
horizontal Maxwell's construction in $\mu$-$Q$ diagram.
This is a process where the chemical potential is used
as the control parameter.
Such a process is more typical in the literature
\cite{Yagi:2005yb}. 
Some other related questions are
about the effect of the density and temperature on the meson spectrum
and on the heavy quark potentials. 
A limitation of this model is the absence of gluon confinement. 
We can also ask whether what we found
in this paper is universal feature of black hole embeddings.
These issues are currently under investigation \cite{work2}.

{\bf  Acknowledgments: }{ \small
The authors want to thank Kazuaki Ohnishi, Mannque Rho 
and Motoi Tachibana for useful discussions.
This work was supported by KOSEF Grant R01-2004-000-10520-0 and
the SRC Program of the KOSEF through the Center for Quantum
Space-time of Sogang University with grant number R11-2005-021.}

\end{document}